\documentclass[10pt,conference]{IEEEtran}
\pdfoutput = 1

\usepackage{algorithmic}
\usepackage{amsmath,amssymb,amsfonts}
\usepackage{cite}
\usepackage{comment}
\usepackage{listings}
\usepackage{lipsum}
\usepackage{graphicx}
\usepackage[colorlinks,urlcolor=black,linkcolor=black,citecolor=black,linktocpage=true]{hyperref}
\usepackage{siunitx}
\usepackage{subfigure}
\usepackage{textcomp}
\usepackage{xcolor}
\usepackage{xspace}

\newcommand{\SWH}{Software Heritage\xspace}

\newcommand{\EMAIL}[1]{\href{mailto:#1}{#1}}
\newcommand{\TODO}[1]{{\color{magenta} \textbf{TODO}\ifthenelse{\equal{#1}{}}{\xspace}{:~}#1 }}
\newcommand{\URL}[1]{\href{https://#1}{#1}}

\newcommand{\SWHFS}{SwhFS\xspace}
\newcommand{\SWHID}[1]{\href{https://archive.softwareheritage.org/#1}{#1}}
\newcommand{\SCREENCASTURL}{\URL{dx.doi.org/10.5281/zenodo.4531411}}
\newcommand{\REPLICATIONURL}{\URL{dx.doi.org/10.5281/zenodo.4535531}}

\lstdefinestyle{shelllog}{
  language=bash,
  morekeywords={
    find,
    grep,
    head,
    jq,
    ls,
    mkdir,
    wc,
    xargs,
  },
  deletekeywords={
    hash,
    history,
  },
  showstringspaces=false,
}
\newcommand{\SWHFSpy}{\lstinline[language=Python]{swh.fuse}\xspace}
\newcommand{\TEXTTT}[1]{\texttt{\small #1}}

\title{The Software Heritage Filesystem (SwhFS):\\
  Integrating Source Code Archival with Development} \author{\IEEEauthorblockN{Thibault Allançon}
  \IEEEauthorblockA{EPITA and Inria\\
    Paris, France\\
    \EMAIL{thibault.allancon@epita.fr}
  }
  \and
  \IEEEauthorblockN{Antoine Pietri}
  \IEEEauthorblockA{Inria\\
    Paris, France\\
    \EMAIL{antoine.pietri@inria.fr}
  }
  \and
  \IEEEauthorblockN{Stefano Zacchiroli}
  \IEEEauthorblockA{Université de Paris and Inria\\
    Paris, France\\
    \EMAIL{zack@irif.fr}
  }
}

\begin{document}
\maketitle

\begin{abstract}
  We introduce the Software Heritage filesystem (SwhFS), a user-space
  filesystem that integrates large-scale open source software archival with
  development workflows.  SwhFS provides a POSIX filesystem view of Software
  Heritage, the largest public archive of software source code and version
  control system (VCS) development history.

  Using SwhFS, developers can quickly ``checkout'' any of the 2 billion commits
  archived by Software Heritage, even after they disappear from their previous
  known location and without incurring the performance cost of repository
  cloning. \SWHFS works across unrelated repositories and different VCS
  technologies. Other source code artifacts archived by Software
  Heritage---individual source code files and trees, releases, and
  branches---can also be accessed using common programming tools and custom
  scripts, as if they were locally available.

  A screencast of SwhFS is available online at \SCREENCASTURL.

\end{abstract}


 \section{Introduction}
\label{sec:intro}

Distributed version control (DVCS) is the state-of-the-art for source code
management in software development. With DVCSs the full development history is
replicated by each developer and code changes are exchanged among peers.

DVCS \emph{per se} does not guarantee \emph{code availability} though. Public
repositories can disappear: they can be made private or moved; they can be
targeted by (potentially bogus) copyright takedown notices; their hosting
platform might shutdown. Specific commits can also disappear from repositories,
e.g., due to their history being rewritten. The commits will still be available
from other copies of the repository, but there is no easy way to find them.

Large-scale source code archival is a crucial part of a solution to this
problem, especially for the vast corpus of free/open source software.
\SWH~\cite{swhipres2017, swhcacm2018} is the largest archive of public code,
having archived at the time of writing almost 10 billion (B) unique source code
files and 2\,B unique commits from more than 150\,M projects spanning multiple
hosting platforms and VCS technologies.

Source code artifacts in the \SWH archive (source code file or tree, commit,
release, etc.) are stored in a global Merkle DAG and identified by
content-based persistent identifiers called SWHIDs~\cite{swhipres2018}, e.g.,
\SWHID{swh:1:rev:9d76c0b163675505d1a901e5fe5249a2c55609bc}. Given a SWHID, one
can browse what's behind it via a Web
UI.\footnote{\url{https://archive.softwareheritage.org/}} SWHIDs for source
code that is not locally available can be obtained in various ways: searching
archived software by origin URL or metadata; finding SWHIDs mentioned in
scientific papers~\cite{biblatex-software}, Wikidata, or software bills of
materials~\cite{stewart2010spdx}; deriving SWHIDs from other VCS references
(e.g., ``9d76c0b163675505d1a901e5fe5249a2c55609bc'' in the previous example is
a Git-compatible commit identifier).

\paragraph*{Contributions and use cases}

We introduce the \emph{Software Heritage Filesystem
  (\SWHFS)},\footnote{\url{https://docs.softwareheritage.org/devel/swh-fuse/}}
a user-space filesystem that gives access to the \SWH archive as a POSIX
filesystem. \SWHFS integrates with development workflows by ensuring code
availability when desired artifacts disappear from their last known location.
Provided a reference to the desired code can be obtained, the corresponding
file, directory, or commit can be used as if it were locally available.

Even for still-available code, \SWHFS offers a convenient way to explore huge
code bases~\cite{msscalar, msvfsforgit} that require significant time to be
fully retrieved over the network, large amounts of disk space, and high I/O
costs when switching branches.

\begin{figure*}
  \centering
  \includegraphics[width=0.9\textwidth]{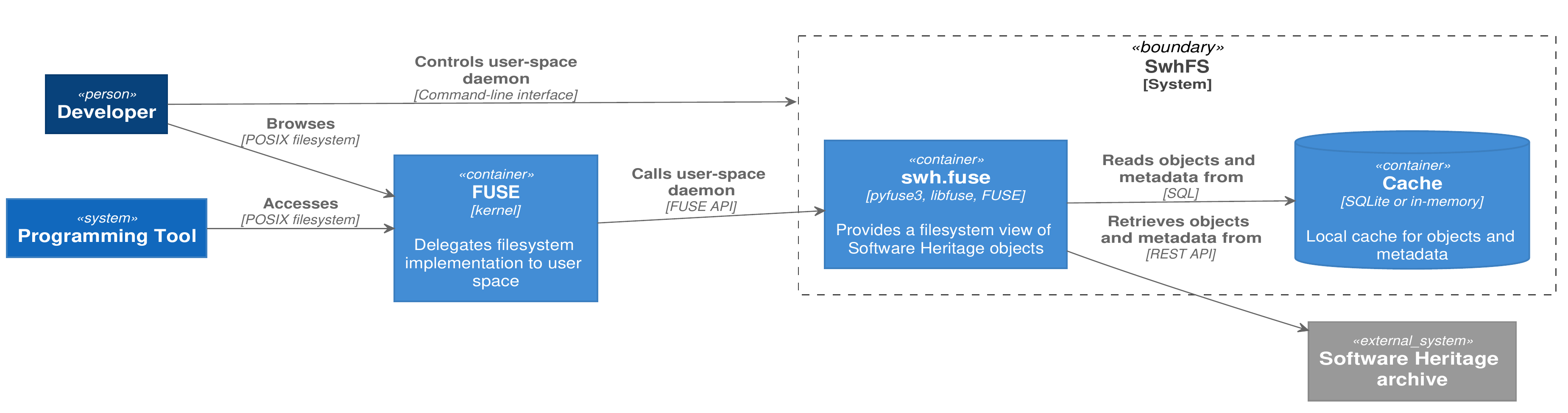}
  \caption{Architecture of the Software Heritage virtual user-space filesystem
    (SwhFS)}
  \label{fig:swhfs-arch}
\end{figure*}

 \section{Related Work}
\label{sec:related}

Other VCS filesystems have been proposed in the past.
RepoFS~\cite{spinellis2019repofs} exposes a Git repository as a
FUSE~\cite{fuse, vangoor2017fuseperf} filesystem, making different trade-offs
than \SWHFS. RepoFS needs a \emph{local} Git repository, while \SWHFS loads it
lazily over the network. This makes \SWHFS more suitable for quick exploration,
and RepoFS more suitable for repository mining.

GitOD~\cite{schroeder2012gitod} and VFS for Git~\cite{msvfsforgit} expose
remote Git repositories as local filesystems, loading them lazily in order to
reduce disk and bandwidth usage for huge code bases, retaining compatibility
with standard Git tools.

GitFS~\cite{presslabs-gitfs} and FigFS~\cite{grant2009figfs} also offer virtual
filesystem interfaces on top of Git. FigFS is now abandoned. GitFS is not, but
focuses on the use case of authoring new commits.

All related works reviewed thus far are Git-specific and have a
single-repository scope. \SWHFS is VCS-agnostic and spans the entire \SWH
archive, giving access to hundreds of million code repositories in a unified
view.

To the best of our knowledge \SWHFS is the first attempt to integrate
large-scale source code archival into development workflows via the filesystem
interface.

 \section{Design}
\label{sec:architecture}

Figure~\ref{fig:swhfs-arch} shows the \SWHFS architecture as a C4
diagram~\cite{brown2018c4}.

\paragraph{Front-end: POSIX filesystem and FUSE}

Users control \SWHFS by starting/stopping its user-space daemon via a
convenient CLI interface. While running, \SWHFS provides a filesystem view of
all the source code artifacts archived by \SWH. This view is exposed as a POSIX
filesystem that can be browsed using file-navigation tools and accessed by
programming tools like text editors, IDEs, compilers, debuggers, custom shell
scripts, etc.

According to the FUSE design~\cite{vangoor2017fuseperf}, the POSIX filesystem
interface seen by \SWHFS clients is implemented by the OS kernel, which
delegates decisions about the content of virtual files and directories to a
user-space daemon, communicating with it via the FUSE API~\cite{fuse}. The
\SWHFS daemon is implemented in the \SWHFSpy Python module, using the
\TEXTTT{pyfuse3}
bindings to the C FUSE API.

\paragraph{Filesystem layout}

The filesystem layout implemented by \SWHFSpy consists of (1) a set of entry
points and (2) filesystem representations of \SWH objects. \emph{Entry points}
are located just below the \SWHFS mount point; most notably \TEXTTT{archive/}
allows browsing the \SWH archive by known SWHIDs while \TEXTTT{origin/} allows
doing so by the URLs of archived projects (see Section~\ref{sec:walkthrough}
for examples).

\emph{Filesystem representations} of \SWH objects depend on the object type:
\begin{itemize}

\item \emph{Source code files} (SWHID type: \TEXTTT{cnt}) are represented as
  regular POSIX files; 

\item \emph{Source code directories} (\TEXTTT{dir} type) as directories whose
  content matches what was archived;

\item \emph{Commits and releases} (\TEXTTT{rev} and \TEXTTT{rel}) as
  directories containing a \TEXTTT{root} sub-directory pointing to the source
  code tree plus auxiliary files containing metadata such as commit message and
  ancestry, timestamps, author information, etc.;

\item \emph{Repository snapshots} (\TEXTTT{snp}) as directories containing one
  entry for each repository branch (\TEXTTT{master}, \TEXTTT{v1.0},
  \TEXTTT{bug-6531}, etc.), each pointing to the filesystem representation of
  the corresponding commit or release.

\end{itemize}

Relationships between accessed archived objects are represented using symbolic
links. For instance, the source tree of a commit object
\TEXTTT{archive/swh:1:rev:.../root} is a symbolic link to a directory object
located under \TEXTTT{archive/}, e.g., \TEXTTT{"->~../../swh:1:dir:..."}. This
approach avoids duplications in the virtual filesystem, reifying on disk the
sharing of Merkle structures.

The walkthrough of Section~\ref{sec:walkthrough} presents additional details
about the \SWHFS filesystem layout. A full layout specification is included
with \SWHFS documentation.

\paragraph{Backend: a \SWH $\leftrightarrow$ FUSE adapter}

\SWHFSpy is an adapter between the \SWH REST
API\footnote{\url{https://archive.softwareheritage.org/api/}} (SWH API) and the
FUSE API. When filesystem entities that represent archived objects are
accessed, \SWHFSpy uses the SWH API to retrieve data and metadata about them.
Obtained results are then used to assemble the expected filesystem layout and
return it to the kernel via the FUSE API. For example, when a source tree
object is listed using \lstinline[language=C]{readdir()}, \SWHFS invokes the
\TEXTTT{/directory} endpoint of the \SWH API and return to the kernel its
directory entries using a lazy iterator; when a source code file is
\lstinline[language=C]{open()}-ed, \TEXTTT{/content/raw} is called and byte
chunks of the retrieved blob are returned to the kernel piecemeal.

As part of adaptation, \SWHFSpy takes care and abstracts over details such as
pagination, encoding issues, inode and file description allocation, etc.

\paragraph{Performance optimizations}

Remote filesystems can be notoriously painful to use, due to network overhead.
To make things worse for \SWHFS, the backend technology stack (REST APIs and
long-term archival storage) was not designed with filesystem-level performances
in mind. In order to make it fast enough for practical use, \SWHFS implements
several optimizations.

\paragraph*{Caching}

Several \emph{on-disk caches} are stored in SQLite databases to avoid repeating
remote API calls.
The \emph{blob cache} stores raw source code file contents. The \emph{metadata
  cache} stores the metadata of any kind of looked up object. Using these two
caches it is possible to navigate any part of the \SWH archive that has been
accessed in the past, even while disconnected from the network.

Merkle properties and the fact that \SWHFS is read-only make cache invalidation
unneeded, because no cached object can ever change. It is also always possible
to purge objects from these caches without introducing inconsistencies.

\emph{In-memory caching} is used for directories, as most filesystems do. The
\emph{direntry cache} maps directories to their entries, so that frequently
accessed directories can be listed without even incurring SQLite query costs.

\paragraph*{Compressed in-memory graph representation}

When accessing commits and release objects, developers often need to quickly
explore their commit history, \emph{à la} \TEXTTT{git} \TEXTTT{log}. Exploring
commit histories via the core SWH API would be too slow for that, as one HTTP
call per commit is needed, times tens or hundreds of thousand commits for large
repositories, with no parallelization opportunity, given that commit
identifiers are discovered incrementally.

To address this issue we built upon
\TEXTTT{swh-graph},\footnote{\url{https://docs.softwareheritage.org/devel/swh-graph/}}
a compressed in-memory representation of the \SWH archive, obtained applying
webgraph compression techniques~\cite{saner-2020-swh-graph}. The compressed
structure of the global VCS graph comprising 17\,B nodes and 200\,B edges
fits in $\approx$100\,GiB of RAM and can be visited with amortized traversal
cost close to a single memory access per traversed edge.

We have extended the SWH API to expose the \TEXTTT{swh-graph} graph traversal
API. For commit objects, \SWHFS invokes it to perform a graph visit of all
reachable commits and then populate several \emph{history summary views} which
are available under the \TEXTTT{history/by-date/}, \TEXTTT{history/by-hash/},
and \TEXTTT{history/by-page/} directories (see Section~\ref{sec:walkthrough}
for details). Even for huge repositories like the Linux kernel, retrieving the
full commit history (as a list of SWHIDs) requires just a few tens of seconds,
including network transfer.

\paragraph*{Asynchronicity}

In order to maximize I/O throughput \SWHFS is implemented in asynchronous
style:
all blocking operations yield control to other coroutines instead of blocking.
Additionally, SWH API calls are delegated to a shared thread pool that can keep
persistent HTTP connections to the remote API backend, rather than establishing
new ones at each call.

 \section{Walkthrough}
\label{sec:walkthrough}

This section shows how to use \SWHFS via concrete examples. A screencast video
is available online at \SCREENCASTURL.

\paragraph{Installation}

\SWHFS is implemented in Python, released under the GPL3 license, and
distributed via PyPI.
It can be installed from there running \lstinline{pip install swh.fuse}.

\SWHFS development happens on the \SWH
forge,\footnote{\url{https://forge.softwareheritage.org/source/swh-fuse/}}
where issues and patches can be submitted.

\paragraph{Setup and teardown}

Like with any filesystems, \SWHFS must be ``mounted'' before use and
``unmounted'' afterwards. Users should first mount the \SWH archive as a whole
and then browse archived objects looking up their SWHIDs below the
\TEXTTT{archive/} entry-point. To mount the Software Heritage archive, use the
\lstinline{swh fs mount} command:
\begin{lstlisting}
$ mkdir swhfs
$ swh fs mount swhfs/  # mount the archive
$ ls -F swhfs/  # list entry points
archive/  cache/  origin/  README
\end{lstlisting}
By default \SWHFS daemonizes in background and logs to syslog; it can be kept
in foreground, logging to the console, by passing \TEXTTT{-f/--foreground} to
\TEXTTT{mount}.

To unmount \SWHFS use \lstinline{swh fs umount PATH}. Note that, since \SWHFS
is a \emph{user-space} filesystem, (un)mounting are not privileged operations,
any user can do it.

The configuration file \TEXTTT{\~{}/.swh/config/global.yml} is read if
present. Its main use case is inserting a per-user authentication token for the
SWH API, which might be needed in case of heavy use to bypass the default API
rate limit.

\paragraph{Source code browsing}

Here is a \SWHFS Hello World:\\
\begin{minipage}{1.0\linewidth}
\begin{lstlisting}
$ cd swhfs/
$ cat archive/swh:1:cnt:c839dea9e8e6f0528b4\
68214348fee8669b305b2
\end{lstlisting}
\vspace{-2ex}
\begin{lstlisting}[language=C]
#include <stdio.h>

int main(void) {
    printf("Hello, World!\n");
}
\end{lstlisting}
\end{minipage}\\
Given the SWHID of a source code file, we can directly access it via the
filesystem. We can do the same with entire source code directories. Here is the
historical Apollo 11 source code, where we can find interesting comments about
the antenna during landing:
\begin{lstlisting}
$ cd archive/swh:1:dir:1fee702c7e6d14395bbf\
5ac3598e73bcbf97b030
$ ls | wc -l
127
$ grep -i antenna THE_LUNAR_LANDING.s | cut -f 5
# IS THE LR ANTENNA IN POSITION 1 YET
# BRANCH IF ANTENNA ALREADY IN POSITION 1
\end{lstlisting}

We can checkout the commit of a more modern codebase, like jQuery, and count
its JavaScript lines of code (SLOC):
\begin{lstlisting}
$ cd archive/swh:1:rev:9d76c0b163675505d1a9\
01e5fe5249a2c55609bc
$ ls -F
history/  meta.json@  parent@  parents/  root@
$ find root/src/ \
    -type f -name '*.js' | xargs cat | wc -l
10136
\end{lstlisting}

\paragraph{Commit history browsing}

\TEXTTT{meta.json} contains complete commit metadata, e.g.:
\begin{lstlisting}
$ jq .author.name,.date,.message meta.json
"Michal Golebiowski-Owczarek"
"2020-03-02T23:02:42+01:00"
"Prevent collision with Object.prototype ..."
\end{lstlisting}

Commit history can be browsed commit-by-commit digging into \TEXTTT{parent(s)/}
directories or, more efficiently, using the history summaries located under
\TEXTTT{history/}:
\begin{lstlisting}
$ ls -f history/by-page/000/ | wc -l
6469
$ ls -f history/by-page/000/ | head -n 2
swh:1:rev:358b769a00c3a09a...
swh:1:rev:4a7fc8544e2020c7...
\end{lstlisting}
The jQuery commit at hand is preceded by \num{6469} commits, which can be
listed in ``\TEXTTT{git log}'' order via the \TEXTTT{by-page} view. The
\TEXTTT{by-hash} and \TEXTTT{by-date} views, inspired by
RepoFS~\cite{spinellis2019repofs}, list commits sharded by commit identifier
and timestamp:
\begin{lstlisting}
$ ls history/by-hash/00/ | head -n 1
swh:1:rev:0018f7700bf8004d...
$ ls -F history/by-date/
2006/  2007/  2008/  ...  2018/  2019/  2020/
$ ls -f history/by-date/2020/01/08/
swh:1:rev:437f389a24a6bef...
$ jq .date history/by-date/2020/01/08/*/meta.json
"2020-01-08T00:35:55+01:00"
\end{lstlisting}
Note that to populate the \TEXTTT{by-date} view, metadata about all commits in
the history are needed. To avoid blocking, metadata are retrieved asynchronously
in the background, populating the view incrementally. The hidden
\TEXTTT{by-date/.status} file provides a progress report and is removed upon
completion.

\paragraph{Repository snapshots and branches}

Snapshot objects keep track of where each branch and release (or ``tag'')
pointed to at archival time. Here is an example with the Unix history
repository~\cite{SpinellisUnix2017}, which uses historical Unix releases as
nested branch names:
\begin{lstlisting}
$ cd archive/swh:1:snp:2ca5d6eff8f04a671c0d\
5b13646cede522c64b7d/refs/heads
$ ls -f | wc -l ; ls -f | grep Bell
40
Bell-32V-Snapshot-Development
Bell-Release
$ cd Bell-Release
$ jq .message,.date meta.json
"Bell 32V release ..."
"1979-05-02T23:26:55-05:00"
$ grep core root/usr/src/games/fortune.c
      printf("Memory fault -- core dumped\n");
\end{lstlisting}
Two of the 40 top-level branches correspond to Bell Labs releases. We can dig
into the UNIX/32V \TEXTTT{fortune} implementation instantly, without having to
clone a 1.6\,GiB repository.

\paragraph{Software origins}

Software can also be explored by the URL it was archived from, using the
\TEXTTT{origin/} entry point:
\begin{lstlisting}
$ cd origin/
$ cd https$ ls
2015-07-09/  2016-02-23/  2016-03-28/  ...
$ ls -F 2015-07-09/
meta.json  snapshot@
\end{lstlisting}
we can see a list of all archival crawls of the Linux kernel repository made by
Software Heritage, and then navigate to the state of the repository as it was
in 2015 (as a snapshot object). Note that one needs to use the exact origin URL
and percent-encode it. To help with that, the companion \texttt{swh web search}
CLI tool is available:
\begin{lstlisting}
$ swh web search "torvalds linux" \
--limit 1 --url-encode | cut -f1
https\end{lstlisting}

 \section{Conclusion}
\label{sec:conclusion}

We introduced \SWHFS, a user-space filesystem that allows browsing source code
artifacts archived by \SWH as a POSIX filesystem. \SWHFS integrates archival of
public code with development workflows, allowing to quickly ``checkout'' any
archived source code file, tree, commit, or repository without incurring the
full repository cloning costs.  \SWHFS works across unrelated repositories and
different version control technologies. \SWHFS gives access to more than 9\,B
source code files and 2\,B commits, archived by \SWH from more than 140\,M
projects, growing daily.

\paragraph*{Future work}

We plan to increase \SWHFS throughput by making the SWH API queryable for
multiple objects at once. Doing so might be enough to address mining software
repository (MSR) use cases, which are currently out of scope for \SWHFS. We are
also studying the feasibility of integration with stock Git tools, making
commands like \TEXTTT{git log} and \TEXTTT{blame} work within \SWHFS mounts.

Exposing in the filesystem layout additional metadata available from
\SWH---like licensing information and project metadata---is a low-hanging fruit
that can provide added value on top of the already exposed VCS information.

\paragraph*{Data availability}

A replication package for this paper is available online at \REPLICATIONURL.

 \paragraph*{Acknowledgments}

The authors would like to thank Roberto Di Cosmo for his \emph{``how about
  FUSE?''} prompt, which eventually led to \SWHFS.


\begin{thebibliography}{10}

\bibitem{fuse}
Filesystem in userspace ({FUSE}).
\newblock \url{https://github.com/libfuse/libfuse}.
\newblock Accessed 2020-09-29.

\bibitem{swhcacm2018}
Jean-Fran\c{c}ois Abramatic, Roberto Di~Cosmo, and Stefano Zacchiroli.
\newblock Building the universal archive of source code.
\newblock {\em Communications of the {ACM}}, 61(10):29--31, September 2018.

\bibitem{saner-2020-swh-graph}
Paolo Boldi, Antoine Pietri, Sebastiano Vigna, and Stefano Zacchiroli.
\newblock Ultra-large-scale repository analysis via graph compression.
\newblock In {\em SANER 2020: The 27th IEEE International Conference on
  Software Analysis, Evolution and Reengineering}. IEEE, 2020.

\bibitem{brown2018c4}
Simon Brown.
\newblock The {C4} model for software architecture.
\newblock \url{https://c4model.com/}, 2018.
\newblock Accessed 2020-11-16.

\bibitem{biblatex-software}
Roberto~Di Cosmo.
\newblock Announcing biblatex-software: software citation made easy.
\newblock {\em {ACM} {SIGSOFT} Softw. Eng. Notes}, 45(4):22--23, 2020.

\bibitem{swhipres2018}
Roberto Di~Cosmo, Morane Gruenpeter, and Stefano Zacchiroli.
\newblock Identifiers for digital objects: the case of software source code
  preservation.
\newblock In {\em {iPRES} 2018: the 15th Intl.~Conference on Digital
  Preservation}, 2018.

\bibitem{swhipres2017}
Roberto Di~Cosmo and Stefano Zacchiroli.
\newblock {Software Heritage}: Why and how to preserve software source code.
\newblock In {\em {iPRES} 2017: the 14th International Conference on Digital
  Preservation}, 2017.

\bibitem{grant2009figfs}
Reilly Grant.
\newblock Filesystem interface for the git version control system-final report.
\newblock Technical report, University of Pennsylvania, 2009.
\newblock \url{https://www.stwing.upenn.edu/~rm/figfs/final.pdf}, accessed
  2020-11-20.

\bibitem{msscalar}
Microsoft.
\newblock Scalar.
\newblock \url{https://github.com/microsoft/scalar}.
\newblock Accessed 2020-09-29.

\bibitem{msvfsforgit}
Microsoft.
\newblock {VFS} for {Git}.
\newblock \url{https://vfsforgit.org/}.
\newblock Accessed 2020-09-29.

\bibitem{presslabs-gitfs}
Presslabs.
\newblock gitfs: Version controlled file system.
\newblock \url{https://github.com/presslabs/gitfs}, 2014.
\newblock Accessed 2020-11-17.

\bibitem{spinellis2019repofs}
Vitalis Salis and Diomidis Spinellis.
\newblock {RepoFS}: File system view of {Git} repositories.
\newblock {\em SoftwareX}, 9:288--292, 2019.

\bibitem{schroeder2012gitod}
Jonatan Schroeder.
\newblock {GitOD}: An on demand distributed file system approach to version
  control.
\newblock In {\em {CTS} 2012: International Conference on Collaboration
  Technologies and Systems}, pages 613--615. {IEEE}, 2012.

\bibitem{SpinellisUnix2017}
Diomidis Spinellis.
\newblock A repository of {Unix} history and evolution.
\newblock {\em Empirical Software Engineering}, 22(3):1372--1404, 2017.

\bibitem{stewart2010spdx}
Kate Stewart, Phil Odence, and Esteban Rockett.
\newblock Software package data exchange ({SPDX}) specification.
\newblock {\em IFOSS L. Rev.}, 2:191, 2010.

\bibitem{vangoor2017fuseperf}
Bharath Kumar~Reddy Vangoor, Vasily Tarasov, and Erez Zadok.
\newblock To {FUSE} or not to {FUSE:} performance of user-space file systems.
\newblock In {\em 15th {USENIX} Conference on File and Storage Technologies,
  {FAST} 2017}, pages 59--72. {USENIX} Association, 2017.

\end{thebibliography}
\end{document}